\documentstyle[prl,twocolumn,aps,epsf]{revtex}
\def\epsfig#1#2#3#4
         {
         \epsfysize=#2 \vbox{ \hglue#3 \epsfbox[#4]{#1} }
         }
\def\epsfigrot#1#2#3#4
         {
         \epsfxsize=#2
         \setbox\rotbox=\hbox to #2{\epsfbox[#4]{#1}}
         \vbox{\hglue#3 \rotl\rotbox}
         }
\newbox\rotbox
\input rotate
\begin{document}
\draft
\title{Form-factors computation of  Friedel oscillations in Luttinger
liquids.}
\author{F. Lesage, H. Saleur$^*$}
\address{Department of physics, University of Southern California,
Los-Angeles, CA 90089-0484.}
\date{\today}
\maketitle

\begin{abstract}
We show how to  analytically determine for $g\leq {1\over 2}$
the ``Friedel oscillations''  of charge density induced by a single
impurity in a 1D Luttinger liquid of spinless electrons.

\end{abstract}
\vskip 0.2cm

\narrowtext

The  general problem of a Luttinger liquid interacting
with an impurity - that may have internal degrees of freedom -   has
attracted constant attention.  A first reason
is the wealth of physical applications:
they include  the anisotropic Kondo model,
the double well problem \cite{Leggett} and
washboard potential problem  of dissipative quantum mechanics,
scattering through an impurity in quantum wires \cite{kanefish},
and tunneling through a point contact in the fractional quantum Hall
effect
\cite{moon}. Another reason is that this general  problem is
integrable,
and therefore the possibility of obtaining exact solutions exists.
Until recently
however, such solutions had been restricted to thermodynamic
properties.
While of crucial experimental interest,
correlation functions and related transport and dynamical
properties had remained inaccessible analytically. Numerical
simulations were quite difficult, and not always conclusive.

Recently,  major progress has been
made. Based on a new basis of massless
quasiparticles  suggested by
integrability, together with a generalization
of the Landauer B\"uttiker approach, DC properties
have been exactly computed, in remarkable agreement
with experimental results \cite{fls}. Using the form-factors approach
\cite{smirnov},\cite{mussardo},
the dynamical properties of currents have also been obtained exactly
\cite{lss},
or, more precisely, in closed forms that have an arbitrary
accuracy all the way from the UV to the IR fixed points .

The method used in \cite{lss}  worked only for currents, ie
for operators with no anomalous dimension. This was a major drawback,
since  many
physical properties
are described by more complicated operators.  We show
in this letter how operators with non trivial
dimension can  be also handled after some appropriate regularization.
As an application, we determine
the  $2k_F$ part of the
charge density profile in a one dimensional Luttinger liquid
away from an  impurity, a problem which has attracted a lot
of interest recently \cite{Grabert},\cite{Schmitteckert}
\cite{aff2}.

We start with the bosonised form of the model. The
hamiltonian takes the form~:
\begin{equation}
H=\int_{-\infty}^\infty dx \ [8\pi g\Pi^2+\frac{1}{8\pi g}
(\partial_x \phi)^2]+
\lambda \cos\phi(0),
\end{equation}
where we have set $v_F=g$.  Then for the  Friedel oscillations,
the charge density operator is just~:
\begin{equation}
\rho(x)=\rho_0+2 \partial_x\phi +
\frac {k_F}{\pi} \cos[2k_Fx+\phi(x)].
\end{equation}
with $\rho_0=\frac{k_F}{\pi}$ the background charge. We
decompose  this system into even and odd basis \cite{lls} by
decomposing $\phi=\phi_L+\phi_R$ and setting~:
\begin{eqnarray}
\varphi^e(x+t)={1\over\sqrt{2}}\left[\phi_L(x,t)+\phi_R(-x,t)\right]
\nonumber\\
\varphi^o(x+t)={1\over\sqrt{2}}\left[\phi_L(x,t)-\phi_R(-x,t)\right]
\end{eqnarray}
Observe that these two field are left movers. We now fold the system
by setting~:
\begin{eqnarray}
&\phi^e_L=\sqrt{2}\varphi^e(x+t),\ x<0\
\phi^e_R=\sqrt{2}\phi^e(-x+t),
x<0\nonumber\\
&\phi^o_L=\sqrt{2}\varphi^e(x+t),\ x<0\
\phi^o_R=-\sqrt{2}\phi^e(-x+t), x<0
\end{eqnarray}
and introduce new fields $\phi^{e,o}=\phi^{e,o}_L+\phi^{e,o}_R$.
The density oscillations now read~:
\begin{equation}
\frac{\langle\rho(x)-\rho_0\rangle}{\rho_0}=\cos(2k_Fx+\eta_F)
\langle \cos\frac{\phi^o(x)}{2}\rangle \langle
\cos\frac{\phi^e(x)}{2}
\rangle,
\end{equation}
with $\eta_F$ the additional phase shift coming from the unitary
transformation to eliminate the forward scattering term.
$\phi^o$ is the odd field with Dirichlet boundary conditions,
at the origin $\phi^o(0)=0$ leading to \cite{gogo}~:
\begin{equation}
\langle \cos\frac{\phi^o}{2}\rangle \propto
\left(\frac{1}{x}\right)^{g/2},
\label{diri}
\end{equation}
and the $\phi^e$ part is computed with the hamiltonian~:
\begin{equation}
\label{hamil}
H^e=\frac{1}{2}\int_{-\infty}^0 dx \
[8\pi g\Pi^{e2}+\frac{1}{8\pi g}(\partial_x\phi^e)^2]+
\lambda \cos\frac{\phi^e(0)}{2}.
\end{equation}
On general grounds, we expect the scaling form~:
\begin{equation}
\langle \cos\frac{\phi^o}{2}\rangle \propto
\left(\frac{1}{x}\right)^{g/2}F(\lambda x^{1-g}),
\label{scalingfct}
\end{equation}
where $F$ is a scaling function to be determined. Note that
even the small $x$ behaviour of this function was no known in
general.

Our approach is based on  the fact that both systems are
integrable. By considering the free boson as a limit
of the sine-Gordon model \cite{massless}, we describe
it  using
a basis of quasi particle states, the quasiparticles
being massless solitons/antisolitons and breathers, with
factorized scattering. The boundary interaction is
then described by a
scattering matrix, which is elastic \cite{ghozamo},\cite{paulK}.

To compute the  correlation functions , it is convenient
to represent the boundary interaction through a
boundary state \cite{ghozamo} $|B\rangle$~:
\begin{equation}
\label{bdstate}
|B\rangle =\exp\left[
\sum_{\epsilon_1,\epsilon_2}\int_{-\infty}^\infty
\frac{d\theta}{2\pi} Z^{*(L)}_{\epsilon_1} (\theta)
 Z^{*(R)}_{\epsilon_2}
(\theta) K^{\epsilon_1\epsilon_2}(\theta_B-\theta)
\right],
\end{equation}
where the $\epsilon_i$'s denote the type of particles
(solitons/anti-solitons or breathers) and the superscript
denotes whether they are left or right movers since they
are massless particles.  Here $\theta_B$ is a scale related
to $\lambda$ encoding the boundary interaction,
$\lambda\rightarrow 0$ corresponds to $\theta_B\rightarrow -\infty$
and $\lambda \rightarrow \infty$ to $\theta_B\rightarrow \infty$.
$K_{\epsilon\epsilon'}$ is related to the reflection matrices
\cite{ghozamo}.  As usual we have used rapidity variables $\theta$
to encode energy and momentum. For solitons
and and antisolitons for instance,   $e=\pm p=\mu e^\theta$, $\mu$ an
arbitrary
energy scale.

The one point function of interest reads then
$\langle 0|\cos\frac{\phi}{2}|B\rangle$. To use (\ref{bdstate}),
we need first the matrix elements of the operator
$\cos\frac{\phi}{2}$
in the quasiparticle basis: these
follow easily from the massive sine-Gordon form-factors
 \cite{smirnov}. Unfortunately, as discussed briefly in \cite{lss},
the
resulting integrals are all IR divergent! This was
not the case for  the current operator, whose form factor
has the naive engineering dimension of an energy, leading to
convergent integrals. Some sort of regularization is needed,
and the correlations of $\cos\frac{\phi}{2}$ with a boundary  had
remained  so far inaccessible. Our purpose
is to show  how to cure this problem.

To explain our strategy, we consider first
the case  $g=1/2$. Here,
the friedel oscillations are simply \cite{lls}
related to the   spin one point function in an
Ising model with  boundary magnetic field. By using
the same approach as the one outlined before,
one finds the following form-factors expansion~:
\begin{eqnarray}
\langle \sigma(x)\rangle=\sum_{n=0}^\infty \frac{1}{n!}
\int_{-\infty}^\infty \prod_{i=1}^n &\left\{
\frac{d\theta_i}{2\pi} \tanh\frac{\theta_B-\theta_i}{2}
e^{-2\mu x e^{\theta_i}}\right\} \nonumber \\
&\prod_{i<j} \left(\tanh\frac{\theta_i-\theta_j}{2}\right)^2.
\end{eqnarray}
The integrals are all divergent at low energies,
when $\theta_i\rightarrow
-\infty$ and  the integrand goes to a constant. Let
 us  then introduce an IR cut-off ( we chose
 $\theta\geq \theta_{min}$ and set $\Lambda\equiv e^{\theta_{min}}$)
and
take the log of the previous expressions (a similar method has been
used in \cite{smirnov} to study the UV limit
of massive correlators. See also , \cite{zamo},\cite{mussardocardy}).
 Ordering this log by
increasing
number of integrations, one can show that each term
diverges as $\ln\Lambda$. Moreover, since
the divergence occurs at very low energy, where the $\tanh$ goes to
unity,
the amplitudes of these $\ln\Lambda$ do {\bf not} depend
on $\theta_B$ (for $\theta_B\neq -\infty$), ie on the boundary
coupling. It is
then easy to get rid of the cut-off: we simply
substract
the log of the IR spin function, ie we substract the same formal
expression with $\theta_B= \infty$. The first two terms of the
resulting expression read~:
\begin{eqnarray}
\ln\frac{\langle \sigma(x)\rangle_{T_B}}{\langle
\sigma(x)\rangle_{IR}}
=\int_{\Lambda}^\infty \frac{du}{2\pi u} e^{-2u x}
\left( \frac{T_B-u}{T_B+u}-1\right)\nonumber \\
+\frac{1}{2}\int_\Lambda^\infty \prod_{i=1}^2 \frac{du_i}{2\pi u_i}
e^{-2\mu  u_i x} \left( \prod_{i=1}^2
\frac{T_B-u_i}{T_B+u_i}-1\right)
\nonumber \\ \times
\left[\left(\frac{u_1-u_2}{u_1+u_2}\right)^2-1\right]+\cdots
\label{sigff}
\end{eqnarray}
where we have set  $\mu=1$, $u_i=e^{\theta_i}$,
$T_B=e^{\theta_B}\propto \lambda^{1/(1-g)}$.

\vbox{
\epsfysize=8cm
\epsfxsize=8cm
\epsffile{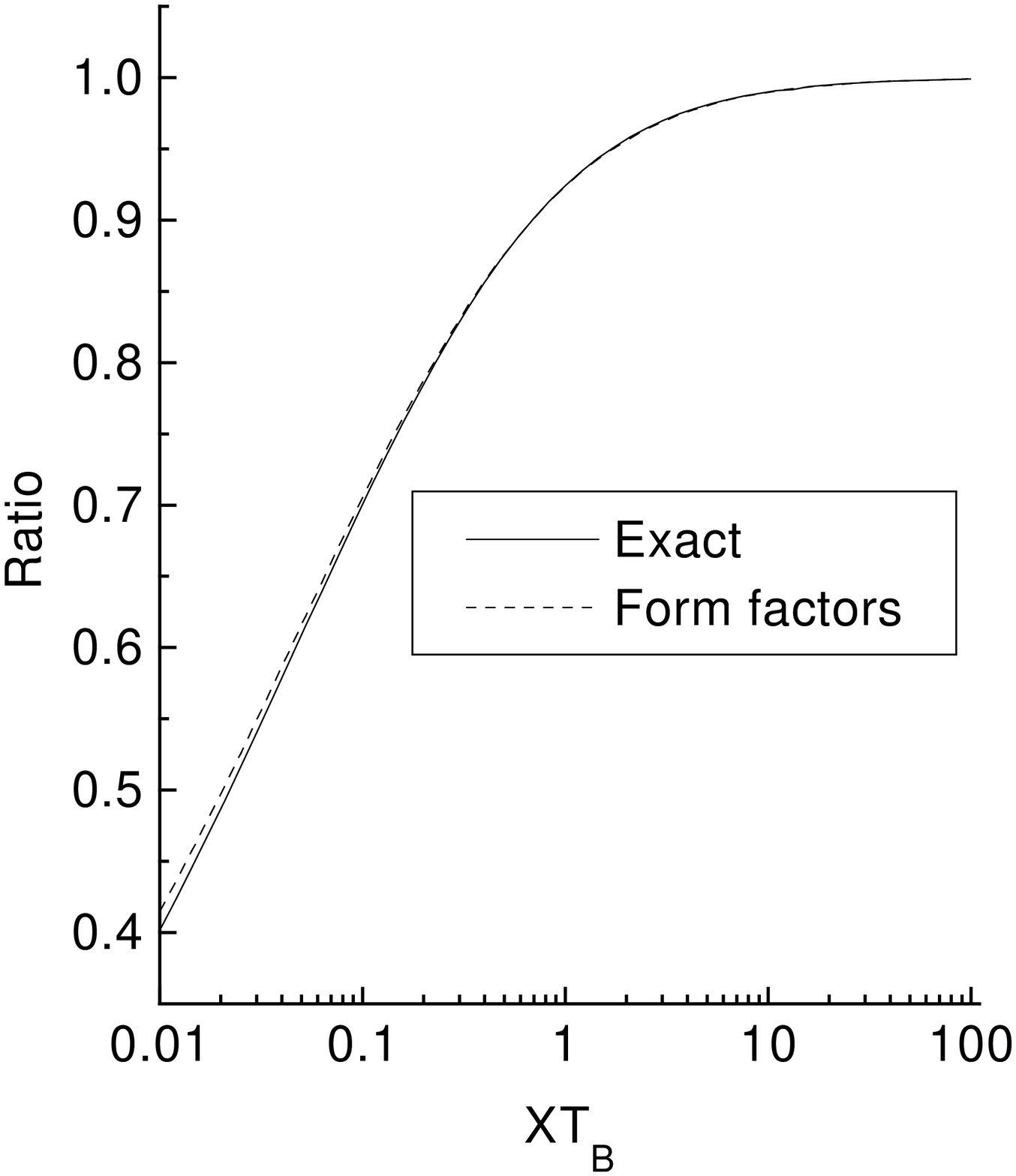}
\begin{figure}
\caption[]{\label{fig1} Accuracy of the finite $T_B$ over the IR
part of the envelope of $\rho(x)$ for $g=1/2$.}
\end{figure}}

Clearly, the integrals are now convergent at low energies, and we can
send $\Lambda$
to zero. Since the IR value of the one point function is easily
determined by other means, $<\sigma(x)>_{IR}\propto x^{-1/8}$
\cite{cardlew},
we can now obtain $<\sigma(x)>_{T_B}$ from (\ref{sigff}). Hence the
procedure involves a {\bf double} regularization. Of course,
there
remains an infinity of terms
to sum over. However, as in the case of current operators, the
convergence
of the form-factors expansion is very quick, and the first few terms
are sufficient to get excellent accuracy all the way from UV to
IR. To illustrate this more precisely, we recall
that  for $g=1/2$ (\ref{sigff}) can be resummed in closed form,
giving rise to~:
\begin{equation}
R_{exact}=
\frac{\langle \sigma(x)\rangle_{T_B}}{\langle \sigma(x)\rangle_{IR}}
=\frac{1}{\sqrt{\pi}} \sqrt{2 xT_B} e^{xT_B} K_0(x T_B).
\label{exact}
\end{equation}
By reexponentianing  the two first terms in (\ref{sigff}),
one gets a ratio differing from (\ref{exact})
by at most $1/100$ for $xT_B\in [0,\infty)$ (see figure 1).

By reexponentiating
the first three terms, acuracy is improved to more than $1/1000$.
Clearly,
the form-factors approach thus provides  analytical expressions
that can be considered as exact for most reasonable purposes.

It is fair to mention however that, at any given order in
(\ref{sigff}), the
exponent controlling the $x\to 0$ behaviour is not exactly
reproduced, as could be seen on a log-log plot.
For instance, the first term is immediately found to produce a
behaviour $R(x)\propto x^{1/\pi}$,
to be compared with the result $R_{exact}(x)\propto x^{1/2}\ln x$.
The
comparison  of the exact result (\ref{exact}) and of (\ref{sigff})
show that the form factors expression has, term by term, the correct
asymptotic expansion
ie the  IR expansion in powers of ${1\over xT_B}$. Adding terms with
more form-factors simply gives a more accurate determination
of the coefficients. This is to be compared with   the
results of \cite{lss} for eg the frequency dependent conductance,
where the form factors expression had the correct functional
dependence both in the UV and in the IR.  This is not to say that our
method
is inefficient in the UV, because we know, at least formally, all the
terms.
In fact, we will show in what follows how (\ref{sigff}) can always be
resummed
in the UV, and that the exponent can be exactly obtained from our
approach too.

The regularization  is the
same for other values of $g$. Here, we discuss
 $g=1/t$ with $t$ integer.  For these values,
the scattering  is diagonal and
the form factors  are rather simple.  To obtain
them, we take the  massless limit of the results
in \cite{smirnov}
and impose  that half of the
quasiparticles  become right movers and half become left movers,
since
the boundary state always involve pairs of right and left moving
particles.  It is in fact easier  to take that limit if we change
basis from the solitons and anti-solitons to $\frac{1}{\sqrt{2}}(
|S>\pm |A>)$.  In that case, the boundary scattering matrix
become diagonal and the isotopic indices always come in pairs.
The reflection matrices in this new basis are given by~:
\begin{eqnarray}
&K_-(\theta)=-e^{i\frac{\pi}{4}(2-t)} \tanh(\frac{(t-1)\theta}{2}
+i\frac{\pi(t-2 )}{4 }) R(i\frac{\pi}{2}-
\theta) \nonumber \\
&K_+(\theta)=e^{i\frac{\pi}{4}(2-t)} R(i\frac{\pi}{2}-\theta)
\end{eqnarray}
with~:
\begin{equation}
R(\theta)=\exp\left( i \int_{-\infty}^\infty \frac{dy}{2y}
\frac{\sin(\frac{2(t-1) y\theta}{\pi})\sinh((t-2)y)}
{\sinh(2y)
\cosh((t-1)y)}\right).
\end{equation}
The breathers reflection matrices are given in \cite{gho}.

The case $g=1/2$ has already been worked out, so let us concentrate
on
$g=1/3$ as an example.  Then, in addition to the soliton and
anti-soliton,
there is also one breather.
The first contribution to the one point function comes  from
the two breathers  form-factor, with one right moving and
one left moving breather. It is given by a constant~:
\begin{equation}
f(\theta,\theta)_{11}^{LR}=c_1,
\end{equation}
and this obviously leads to IR divergences.
Other contributions come from $2n$ breathers form-factors, and $4n$
solitons form-factors
The whole expression can be controlled as for $g=1/2$,
by taking the log, and factoring out the IR part.  Setting
$c(x)= \cos\frac{\phi(x)}{2}$, we
organize the sum as follows~:
\begin{equation}\label{tata}
\ln\frac{\langle c(x)\rangle_{T_B}}{\langle c(x)\rangle_{IR}}=\ln
R^{(2)}+
\ln R^{(4)}+
\cdots
\end{equation}
with the subscript denoting  the number of intermediate excitations.

Then, using the explicit expressions for $g=1/3$ we find~:
\begin{equation}
\ln R^{(2)}=2 c_1 e^{2\sqrt{2}T_B x} Ei(-2\sqrt{2}T_B x),
\end{equation}
where $E_i$ is the standard exponential integral. The next
term $\ln R^{(4)}$ is a bit bulky to be written here,
but it is very easy to obtain - similar expressions
have been explicitely given in \cite{lss}.This
is all what is needed for an accuracy better than 1 percent.
In figure 1 we present the results of the ratio at
$g=1/2,1/3,1/4$ for the Friedel oscillations.
It should be noted that this ratio is just the pinning function
of reference \cite{Grabert} and our results agree well qualitatively
with the results found there.

As mentioned before, the deep UV behaviour is a little more difficult
to obtain: the accuracy is good because the ratio goes to zero
anyway,
but the numerical evaluation of the power law would not be too
accurate
with the number of terms we consider. Fortunately,
the full form-factors expansion allows the analytic determination
of this exponent. First, observe
for instance that in (\ref{sigff}) the integrals converge for all
$T_B\neq 0$,
but strictly at $T_B=0$, they do not. To get the dependence of
$\langle c(x)\rangle$
as $T_B\to 0$, we will consider, the logarithm of another ratio,
$\ln\frac{\langle
c(x)\rangle_{T_B}}{\langle c(x')\rangle_{T_B}}$,
where $x$ and $x'$ are two arbitrary coordinates. For this ratio,
even at $T_B=0$,
the integrals are convergent. But $T_B=0$ is the UV fixed point, with
Neumann boundary conditions. While the one point function $\langle
c(x)\rangle_{UV}$
vanishes, the ratio of two such one point functions is well defined,
and can be computed by putting an IR cut-off (a finite system). One
finds
that it goes as $(x/x')^{g/2}$. By regularity as $T_B\to 0$, the same
is true
for the ratio close to $T_B=0$, and thus one has
\begin{equation}
\langle c(x)\rangle \propto (xT_B)^{g/2}, x (T_B)\to 0
\end{equation}
This shows that the universal scaling function in (\ref{scalingfct})
behaves as
$F(y)\propto y^{g\over 1-g}$ for $g<{1\over 2}$. This exponent can
actually be obtained by perturbation theory. Indeed, the first term
in the perturbative expansion of  $\langle c(x)\rangle$ is
\begin{equation}
\lambda x^{g/2}\int_{-\infty}^\infty {dy\over (x^2+y^2)^g}
\end{equation}
For $g<1/2$, this integral diverges in the IR. To regulate it, we
need
to put a new cut-off: since there is no other length scale in the
problem, this
can be nothing but $1/T_B$. Changing variables, the leading behaviour
is $x^{g/2}T_B^g\propto x^{g/2}\lambda^{g/1-g}$, in agreement
with the previous discussion. The exponent   coincides
with the result of the self-consistent harmonic approximation
\cite{Grabert};
but it is important to stress that   the latter is valid only for
$g<<1$.

\vbox{
\epsfysize=8cm
\epsfxsize=8cm
\epsffile{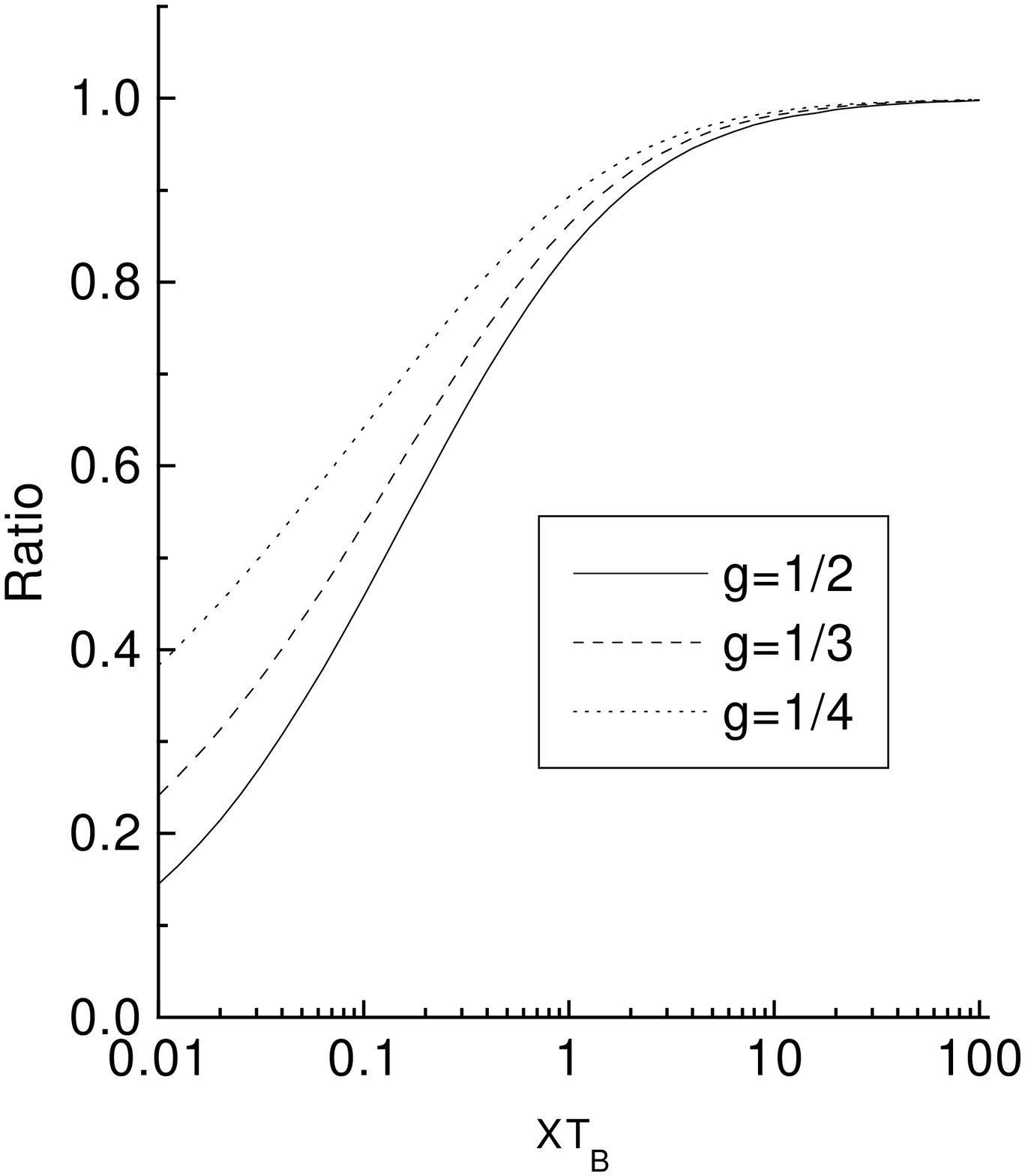}
\begin{figure}
\caption[]{\label{fig2} Ratio of the finite $T_B$ over the IR
part of the envelope of $\rho(x)$.}
\end{figure}}

The function $F(y)$ behaves a $y \ln y$ for $g=1/2$. For
$g>1/2$, its
behaviour is simply $F(y)\propto y$, as can be easily shown
since the perturbative approach is now convergent.
As we approach $g=1/2$ this exponent seems to become asymptotic
and is more difficult to get numerically \cite{Grabert}.

The method presented here is very successful to get
analytical results for $g\leq 1/2$ -  although we limited ourselves
to $g=1/t$ with $t$ integer,
all values of $g<1/2$ are accessible, but
computations are more  complicated since the bulk scattering is non
diagonal.
The method should be generalizable to other problems, in particular
the  determination of the screening cloud in the anisotropic Kondo
model \cite{affleck},
as will be reported elsewhere.
The region $g>1/2$  presents additional
difficulties, unresolved for the moment - in particular,  the
massless
limit of the form factors does not seem to be  meaningful.  Of course
the case $g=1$ can be solved by fermionization \cite{Grabert}. In our
approach,
this point is non trivial
because of  the folding. This folding however is
necessary for any value $g\neq 1$: except at $g=1$, the problem on
the
whole line would not be integrable.

{\bf Ackowledgements:} We thank R. Egger and F. Smirnov
for useful discusions.
This work was supported by the Packard Foundation, the NSF
(through the National Investigator Program) and the DOE.
F.L. was also partially supported by a Canadian NSERC
postdoctoral Fellowship.

 \end{document}